# Gemini Planet Imager observational calibrations V: astrometry and distortion

Quinn M. Konopacky[a], Sandrine J. Thomas[b], Bruce A. Macintosh[c,d], Daren Dillon[e], Naru Sadakuni[e,f], Jérôme Maire[a], Michael Fitzgerald[g], Sasha Hinkley[h], Paul Kalas[i], Thomas Esposito[g], Christian Marois[j], Patrick J. Ingraham[c], Franck Marchis[k], Marshall D. Perrin[l], James R. Graham[i], Jason J. Wang[i], Robert J. De Rosa[m,o], Katie Morzinski[p], Laurent Pueyo[l], Jeffrey K. Chilcote[g], James E. Larkin[g], Daniel Fabrycky[q], Stephen J. Goodsell[f], B.R. Oppenheimer[r], Jenny Patience[m], Leslie Saddlemeyer[j], Anand Sivaramakrishnan[l]

[a]Dunlap Institute for Astronomy and Astrophysics, University of Toronto, 50 St. George St., Toronto, ON M5S 3H4, Canada; [b]NASA Ames Research Center, Moffett Field, Mountain View, CA 94035, USA; [c]Department of Physics, Stanford University, 382 Via Pueblo Mall, Stanford, CA 95305-4060, USA; [d]Lawrence Livermore National Laboratory, 7000 East Avenue, Livermore, CA, 94551, USA; [e]University of California Observatories, UC Santa Cruz, 1156 High Street, Santa Cruz, CA, 95064, USA; [f]Gemini Observatory, Casilla 603, La Serena, Chile; [g]Department of Physics and Astronomy, UCLA, 450 Portola Plaza, Los Angeles, CA 90095, USA; [h]Department of Astronomy, California Institute of Technology, 1200 East California Boulevard, Pasadena, CA 91125, USA; [i]Department of Astronomy, UC Berkeley, Berkeley, CA 94720, USA; [j]National Research Council Canada, Dominion Astrophysical Observatory, 5071 West Saanich Road, Victoria, BC V9E 2E7, Canada; [k]SETI Institute; [l]Space Telescope Science Institute, 3700 San Martin Drive, Baltimore, MD 21218, USA; [m]School of Earth and Space Exploration, Arizona State University, P.O. Box 871404, Tempe, AZ 85287, USA; [o]School of Physics, College of Engineering, Mathematics and Physical Sciences, University of Exeter, Stocker Road, Exeter, EX4 4QL, UK ;[p]Steward Observatory, University of Arizona, 933 North Cherry Avenue, Tucson, AZ 85721, USA; [q]Department of Astronomy and Astrophysics, University of Chicago, 5640 South Ellis Avenue, Chicago, IL 60637, USA; [r]Astrophysics Department, American Museum of Natural History, Central Park West at 79th Street, New York, NY 10024, USA

## ABSTRACT

We present the results of both laboratory and on sky astrometric characterization of the Gemini Planet Imager (GPI). This characterization includes measurement of the pixel scale[*] of the integral field spectrograph (IFS), the position of the detector with respect to north, and optical distortion. Two of these three quantities (pixel scale and distortion) were measured in the laboratory using two transparent grids of spots, one with a square pattern and the other with a random pattern. The pixel scale in the laboratory was also estimate using small movements of the artificial star unit (ASU) in the GPI adaptive optics system. On sky, the pixel scale and the north angle are determined using a number of known binary or multiple systems and Solar System objects, a subsample of which had concurrent measurements at Keck Observatory. Our current estimate of the GPI pixel scale is 14.14 ± 0.01 millarcseconds/pixel, and the north angle is -1.00 ± 0.03°. Distortion is shown to be small, with an average positional residual of 0.26 pixels over the field of view, and is corrected using a 5$^{th}$ order polynomial. We also present results from Monte Carlo simulations of the GPI Exoplanet Survey (GPIES) assuming GPI achieves ~1 milliarcsecond relative astrometric precision. We find that with this precision, we will be able to constrain the eccentricities of all detected planets, and possibly determine the underlying eccentricity distribution of widely separated Jovians.

---

[*]The measurements of pixel scale in this work refer to the rectified scale in the reconstructed data cubes, or the lenslet array pixel scale.

## 1. INTRODUCTION

Exoplanet discovery by direct imaging probes a different region of parameter space that previous detection methods – that of widely separated, Jovian mass objects. With the latest generation of exoplanet imagers now coming online[1,2], the distribution of directly imaged planets (separations of ~10-1000 AU) will begin overlapping with objects discovered via the Doppler method (separations of ~0.01-10 AU), the properties of which have been generally well characterized.[3] A key outstanding question is whether the two populations of exoplanets represent a continuous distribution of objects that form in the same manner, or are two distinct sets of objects. A possible probe of this question is the orbital characteristics of the planets. If, for instance, the populations have similar eccentricity distributions, one could speculate that they stem from the same parent population. Eccentricities also allow for possible constraints of early dynamical evolution, with high eccentricities for widely separated planets implying scattering events and additional planetary bodies in the system.[4]

For planets discovered via direct imaging, dynamical constraints can most easily be made by astrometric monitoring of their position with respect to the host star. Because the orbital periods are long, high precision astrometry is necessary to achieve any constraints in a reasonable time frame. Thus, instruments designed for high contrast imaging must be able to achieve a relative astrometric precision of less than a few milliarcseconds (mas), with higher precision over multiple epochs yielding tighter constraints on orbital parameter distributions.

We present the current results and status of the astrometric characterization and calibration of the newly commissioned Gemini Planet Imager (GPI).[1] The original specifications for GPI called for a relative astrometric precision of 3 mas, with a goal of 1.3 mas. We aim to push even further and achieve 1 mas relative precision per epoch. There are multiple aspects of the calibration effort, as achieving this level of precision is not straightforward and complicated by the science camera being an IFS rather than an imager. The primary star is generally hidden behind the coronagraph occulting spot, presenting a challenge for relative astrometry of any companions. This complication led to the introduction of a diffractive grid in the pupil plane, which generates four reference astrometric satellite spots.[5,6] An analysis of the stability of these spots is presented by Wang et al. [7]. Here, we discuss the measurement of the IFS pixel scale (Section 2) and the orientation of the instrument with respect to the celestial coordinate system (Section 3). We also describe the characterization of the optical distortion in the IFS and our model for correcting it (Section 4). Finally, we discuss the possible eccentricity constraints on the detected planet population we can expect during the GPI Exoplanet Survey (GPIES) if we reach an astrometric precision of 1 mas (Section 5).

## 2. PIXEL SCALE

### 2.1 Laboratory Measurements

A number of laboratory experiments were performed to measure the GPI IFS pixel scale. The first were performed at the University of California, Los Angeles, prior to integration of the IFS and are described in Chilcote et al. [8]. All subsequent tests were conducted after integration of the IFS at the Uinversity of California, Santa Cruz.

The second test, performed on 27 February 2013, utilized artificial star unit (ASU) in the adaptive optics system.[9] The ASU is on a movable stage, which allowed us to combine a series of moves with the known pixel scale internal to GPI (1.610"/mm) to derive the pixel scale. Images were obtained of the ASU on the IFS in *H*-band spectral mode. Significant attenuation of the ASU and an exposure time of 1.5 seconds were used to avoid saturation. Images were taken at twelve different positions. A 3x3 grid was first made using separations in *x* and *y* of 0.5 mm = 0.805". The last three positions were taken at semi-random locations within the interior of the 3x3 grid in order to obtain "off-grid" points.

The data were reduced and transformed into a data cube using the GPI reduction pipeline.[10] We first verified that the centroid of the ASU spot in every data cube was consistent in every wavelength channel. We found that the centroid varied by $\leq 0.03$ pixels. For the primary pixel scale analysis, we therefore elected to focus on channel 18 (1.636 μm), which is approximately midway through the cube.

We measured the centroid of the ASU in each of the twelve IFS frames using the algorithm, which performs point spread function (PSF) fitting provided an input PSF is given.[11] For an input PSF, we used a theoretical PSF generated for the GPI data simulator.[12] StarFinder assigns a correlation value to each source it detects. In the case of the ASU spot, all correlations were above 90%.

Once the centroid in pixels was determined for each frame, we did a pairwise comparison, determining the separation between all independent positions. Figure 1 shows all vectors that went into the final determination of pixel scale in this test. With separation measurements in both pixels and millimeters, we were able to derive a pixel scale. The average of all measurements yields a pixel scale of 14.2 mas/pixel, with a standard deviation of 0.1 mas/pixel.

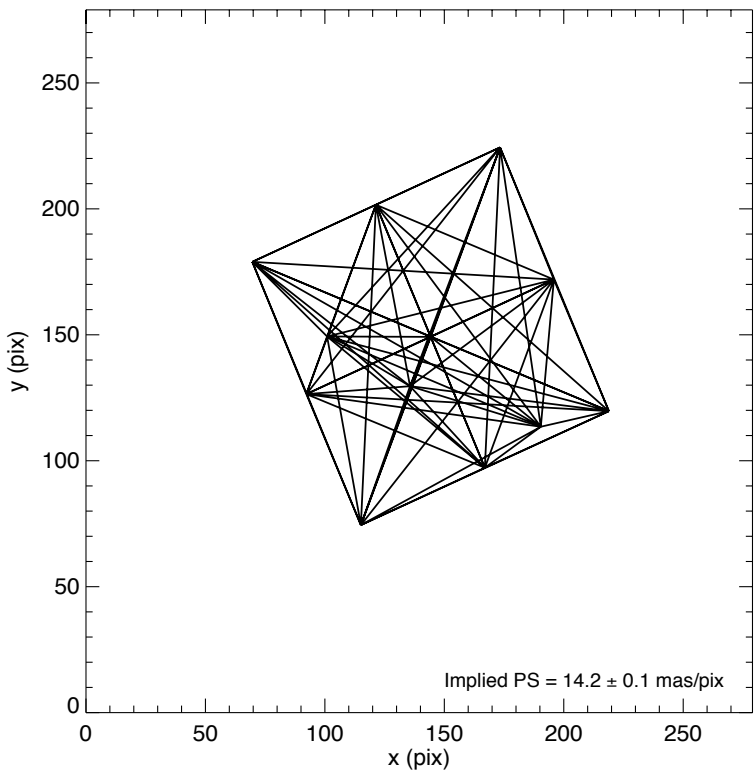

**Figure 1:** Vectors used to determine the pixel scale (shown in pixel units). All independent positions of the ASU were compared once to every other position to yield 66 total separation measurements.

A similar test procedure was used to verify that the pixel scale did not significantly very with temperature or instrument orientation. When GPI was mounted on the flexure rig[13], we rotated it through five different zenith angles – 0, 16, 31, 50, and 80 degrees. At each location, we obtained the data at the same 12 ASU locations. We then reduced and analyzed the data in the same method, and determined the implied pixel scale for each orientation. We found that the maximum change in pixel scale was 0.06%. For temperature verification, we compared our original measurements, which were obtained when GPI was at room temperature (~20° C) to measurements when GPI was in the cold room at an ambient temperature of 0° C. We found that the maximum change in pixel scale with temperature was 0.4%, with all measurements statistically consistent with no change in plate scale.

An additional pixel scale test, also performed on 27 February 2013, utilized a custom-designed pinhole grid consisting of 20x20 square transparent spots on a dark background, printed in metal on a 1" round glass substrate. It is back-illuminated by placing it at an intermediate focus in the GPI telescope simulator downstream from the integrating sphere.[9] IFS images were taken in *H*-band spectral mode. The data were reduced and transformed into a data cube using the GPI data reduction pipeline.[10] An example of the data is shown in Figure 2.

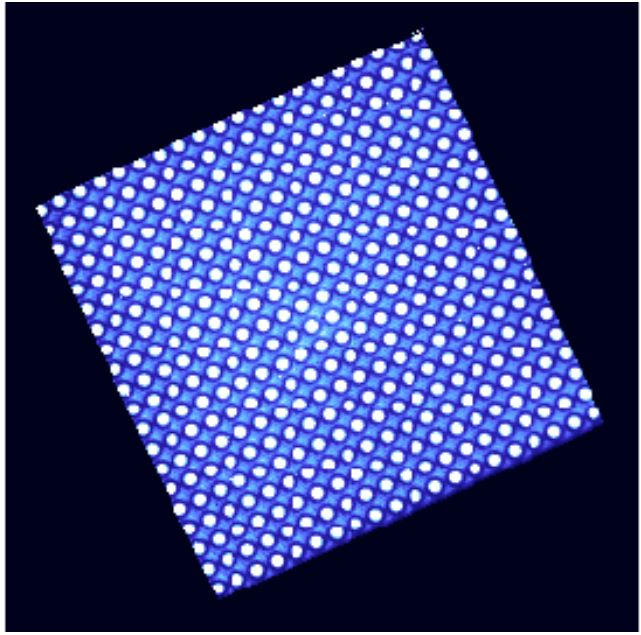

**Figure 2:** Combined image of the illuminated pinhole grid (~1.64 μm). The 20x20 square grid of spots fills the entire reduced FOV.

We again measured the spot centroids using StarFinder and a theoretical PSF.[11] When the grid was designed, we used an estimate for the pixel scale of the telescope simulator to calculate a hole spacing such that the spots would appear separated by 10 pixels. This spacing was calculated to be 65.7 μm (with a spot size of 5 μm). However, there was some uncertainty in both the telescope simulator pixel scale and the precision of the hole spacing. It was immediately apparent when looking at the centroids of the spots that the average spot separation was slightly less than 10 pixels. In order to determine the true spot separation on the detector, we fit for the best single separation value that matched our data. This separation was found to be 9.756 pixels.

Using 9.756 pixels as the separation and combining that with the true separation of 65.7 μm and the estimated telescope simulator pixel scale of 2.13"/mm gives a pixel scale of 14.3 mas/pixel. We derived an uncertainty of 0.15 mas/pixel on this value, based on an estimate of 0.1 pixels uncertainty in the spot separation and 1.5 μm uncertainty in physical hole placement on the grid. We used the pixel scale from the pinhole grid to determine the IFS reconstructed field of view (FOV). This was measured to be 2.67"x2.73". The slight difference in size stems from some spectra generated from lenslets near the edge of the array falling of the edge of the detector, and thus not being extracted. This happens preferentially at the top and bottom edges of the field, where the spectral dispersion direction is orthogonal to the edge of the detector, as opposed to the sides, where the dispersion direction is nearly parallel to the edge of the detector.

## 2.2 On-sky Measurements of Multiple Systems

On-sky measurements of the IFS pixel scale were conducted over multiple observing runs in 2013 November, 2014 March, and 2014 May. In general, the procedure utilized was to observe binary or multiple stars with known separations in arcseconds to translate into a pixel scale. Several types of multiple systems were used for this purpose.

The ideal target for these measurements is a binary or multiple system observed at a near-concurrent time with another well-calibrated instrument. One of the best astrometrically calibrated ground-based cameras is the AO fed camera NIRC2 on Keck II.[14] Five binaries were observed using NIRC2 on three separate nights: 13 March 2014 (PI M. Fitzgerald), 13 May 2014, and 21 May 2014 (PI S. Hinkley). These binaries were identified from a number of catalogs, and include several high contrast binary targets.[15-20] Table 1 summarizes the targets observed with NIRC2, listing the filter and integration time for each source. Data were reduced using standard techniques of dark and background subtraction, bad pixel correction, and flatfield division. They were also adjusted for the known distortion in NIRC2.[14]

Centroids were measured using StarFinder (Diolatti et al. 2000). Pixel positions were converted to separations in arcseconds using the pixel scale given in Yelda et al. [14]. The NIRC2-measured separations and position angles are Table 1.

Table 1. Keck – NIRC2 log of binary star observations and results

| Object | Date Observed | Filter | Total Exp. Time (sec) | Separation (") | Position Angle (deg) |
|---|---|---|---|---|---|
| HIP 43947B | 13 March 2014 | K' | 4.0 | 0.425 ± 0.001 | 260.82 ± 0.04 |
| HIP 44804A | 13 March 2014 | K' | 12.5 | 0.4553 ± 0.0007 | 306.32 ± 0.04 |
| HIP 80628 | 13 March 2014 | Brγ | 0.724 | 0.8591 ± 0.0003 | 43.62 ± 0.02 |
| HIP 85667 | 13 May 2014 | Brγ | 0.477 | 0.788 ± 0.002 | 147.54 ± 0.03 |
| HIP 42074 | 21 May 2014 | K' | 10 | 1.549 ± 0.002 | 220.82 ± 0.05 |

In a number of cases, we observed with GPI sources with no concurrent measurements from other instruments. Some of the earliest GPI observations made were of $\theta^1$ Ori B, which is a quadruple system (B1, B2, B3, and B4) in the Orion Nebula Cluster. The $\theta^1$ Ori B system was analyzed extensively in Close et al. [21] using MagAO on Magellan. By combining new astrometric measurements with those previously obtained over a 15 year time baseline, they were able to determine that the separations of components B1 and B4 and B2 and B3 change only marginally each year - the primary orbital motion seems to affect mostly the positional angle. As such, we were able to use the Close et al. [21] predicted change in separation as a function of time to predict the separations of these objects.

Estimates of the true on-sky separation for five binaries were determined using either orbital ephemeris determination or line fitting to previous astrometric measurements. Table 2 lists the sources in this category and their calculated separations.

Table 2. Binary and multiple system separation estimates from orbits or linear fits

| Object | Estimate from Orbit (") | Estimate from Linear Fit (") | Astrometry Reference |
|---|---|---|---|
| $\theta^1$ Ori B1+B4 | 0.6189 ± 0.0009 | n/a | Close et al. [21] |
| $\theta^1$ Ori B2+B3 | 0.1160 ± 0.0002 | n/a | Close et al. [21] |
| HIP 70931 | 0.72 ± 0.05 | 0.75 ± 0.08 | De Rosa (private comm.) |
| PZ Tel | 0.463 ± 0.006 | n/a | Mugrauer et al. [22] |
| HD 114174 | 0.63 ± 0.03 | 0.659 ± 0.012 | Crepp et al. [21] |

Observations obtained with GPI are summarized in Table 3. Six binaries were observed in coronagraphic mode, while two were observed in unocculted mode. In general, unocculted observations are preferable, as they avoid uncertainties associated with the satellite spots, but are only possible if the primary's brightness is sufficiently faint to avoid saturation.[7] Raw data were converted to cubes using the GPI data reduction pipeline. Images were corrected using the pre-ship estimate of distortion in the IFS (see Section 4). Centroids for both satellite spots and stars were measured by either fitting a Gaussian or with StarFinder using an idealized PSF. If the satellite spots were used, the centroid of the primary was determined by finding the intersection of lines connecting diagonal spots.[7] Separations were then determined, and the implied pixel scale was found using the estimates in Tables 1 and 2. Table 3 gives the final pixel scale and uncertainty for each source. In general the variation in measurement uncertainty for the pixel separation is due to variation in image quality due to the conditions when the data were obtained.

Table 3. GPI log of astrometric calibrator observations and results

| Object | Date Observed | Filter | Mode | Separation (pix) | Implied Pixel Scale (mas/pix) |
|---|---|---|---|---|---|
| θ¹ Ori B1+B4 | 2013 November 13 | H | Coronagraph | 43.47 ± 0.57 | 14.24 ± 0.19 |
| θ¹ Ori B2+B3 | 2013 November 13 | H | Coronagraph | 8.07 ± 0.19 | 14.37 ± 0.34 |
| " | 2014 March 25 | H | Coronagraph | 8.291 ± 0.001 | 13.99 ± 0.04 |
| HIP 43947B | 2014 May 14 | H | Unocculted | 29.962 ± 0.006 | 14.19 ± 0.03 |
| HIP 44804A | 2014 March 23 | K1 | Unocculted | 32.1449 ± 5.1x10$^{-5}$ | 14.17 ± 0.03 |
| " | 2014 May 14 | H | Unocculted | 32.23 ± 0.12 | 14.13 ± 0.06 |
| HIP 70931 | 2014 March 24 | H | Coronagraph | 52.27 ± 0.31 | 14.35 ± 1.53 |
| " | 2014 May 11 | H | Coronagraph | 52.26 ± 0.04 | 14.35 ± 1.53 |
| " | 2014 May 13 | H | Coronagraph | 52.08 ± 0.12 | 14.40 ± 1.54 |
| HIP 80628 | 2014 March 23 | K1 | Coronagraph | 60.80 ± 0.12 | 14.13 ± 0.03 |
| " | 2014 May 11 | K1 | Coronagraph | 60.81 ± 0.06 | 14.13 ± 0.02 |
| HD 114174 | 2014 March 23 | H | Coronagraph | 46.85 ± 0.02 | 14.07 ± 0.26 |
| PZ Tel | 2014 May 11 | H | Coronagraph | 32.34 ± 0.24 | 14.31 ± 0.21 |

Combining all the measurements above gives a pixel scale value of 14.14 ± 0.01 mas/pix. This value is consistent to within the uncertainties of the combination of the laboratory measurements (14.23 ± 0.08 mas/pix), but at the low end of our expectations. However, the number is well within the specifications for the IFS of 14 ± 0.5 mas/pix.[8,23] To provide a feel for the overall pixel scale stability of the instrument, Figure 3 shows all measurements as a function of the date they were taken.

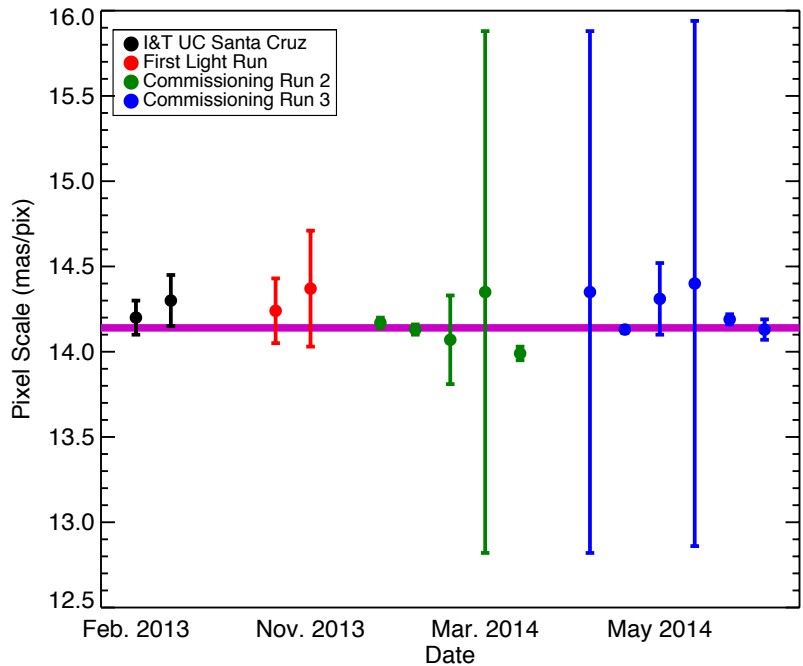

**Figure 3:** All independent pixel scale measurements as a function of the epoch they were obtained. On-sky measurements were obtained during three GPI commissioning runs. The purple line marks the weighted average pixel scale for the IFS, 14.14 ± 0.01 mas/pix. The scale has generally been stable and consistent between observing runs.

### 2.3 On-sky Measurements of Solar System Bodies

Solar system bodies such as large planets and their large satellites can also be used to estimate the pixel scale of detector. Solar system bodies visited and inspected closely by space missions, have their size and shape very well constrained. Europa, satellite of Jupiter was observed on 18 November 2013 in K1 band, and direct mode (no coronagraph, no apodizer). Eleven data cubes with a true integration time of 8.73 s were processed using the data reduction pipeline.[10] Ephemeris calculations provided by JPL & IMCCE indicate that the satellite was bright (V~5.6) with an apparent diameter of 0.9508" and observed with a phase angle of 8.9°, leading to an illumination fraction of 99.4%.

Europa was not dithered on the set of observations. We estimated the residual jittering to be 0.12 pixel on average by fitting the disk with an ellipsoid and measuring its center for each frame. Figure 4 shows the astrometric position of Europa (labeled from E0 to E10), plus the residual jittering measured on seven frames of HD1160 (labeled P0 to P5), a A0 V=7.1 star observed in K1 direct imaging mode on the same night. Because the residual jittering on the "PSF" stars is slightly higher (0.31 pixel in average), we concluded that no residual jittering due to the extended angular size of Europa was introduced. Consequently, it is likely that the angular resolution on these images of Europa is in fact very close to the angular resolution that we measured on the "PSF", so 4.55 ± 0.07 pixels. More interestingly, these images can be used directly to estimate the pixel scale of GPI detector.

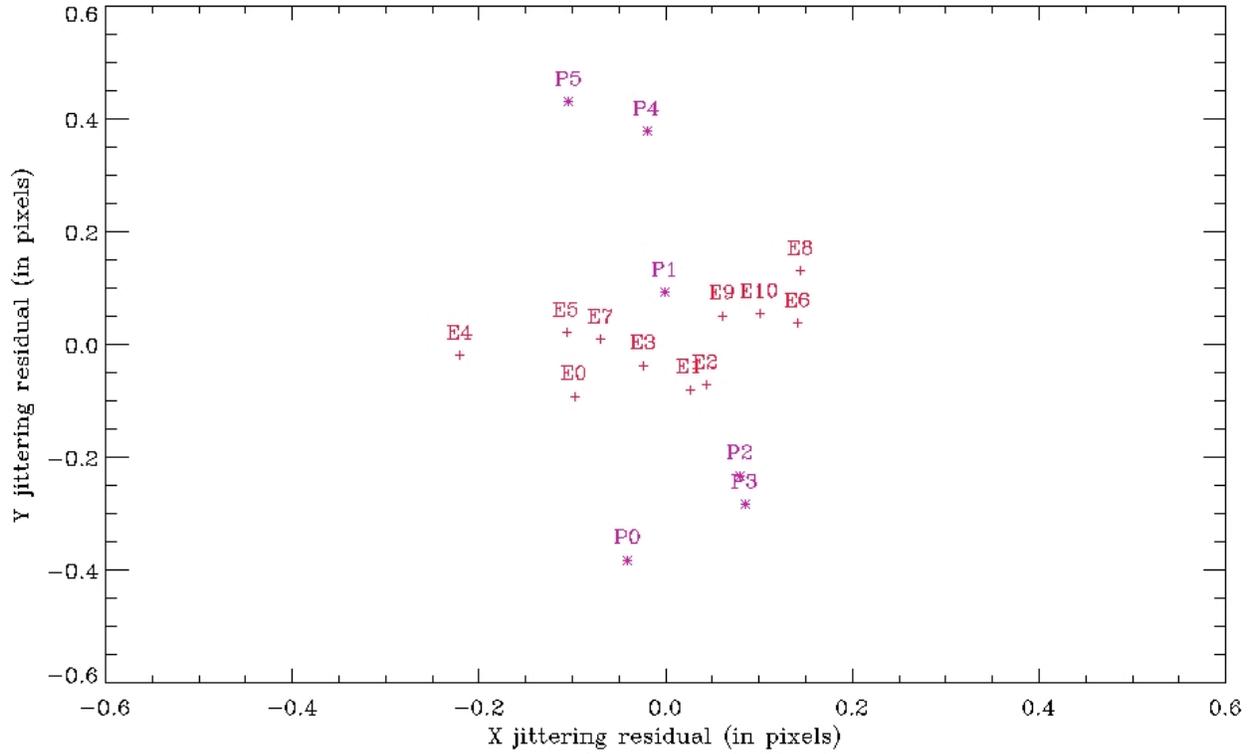

**Figure 4**: Astrometric positions of Europa (E0 to E10) compared to the astrometric position of a star (labeled P0 to P5) observed shortly before Europa. The residual on the star is larger than the residual on Europa implying that the resolved shape of Europa (D~0.95") does not introduce any residual smearing.

We estimated the pixel scale on final assembled cubes shown in Figure 5 by fitting the observations of Europa with a sphere illuminated under the same geometry than Europa (Sub-Solar point and Sub-Earth point were extracted from the ephemeris) and degraded after convolving it by a Gaussian PSF with a FWHM of 4.545 pixels. We generated a grid of solutions varying the Minnaert coefficient, which defines the center-to-limb profile from 1.1 to 1.5 (steps of 0.02) and the pixel scale from 13 to 15 mas (steps of 0.04). The best fit is obtained with a pixel scale estimated to be 14.21 ± 0.21 mas/pixel. This measurement is in agreement with binary star and laboratory measurements presented in Sections 2.1 and 2.2. The error on the measurement may be reduced in the future by introducing an albedo map of Europa in the fitting process to account for the variation on the surface of the satellite.

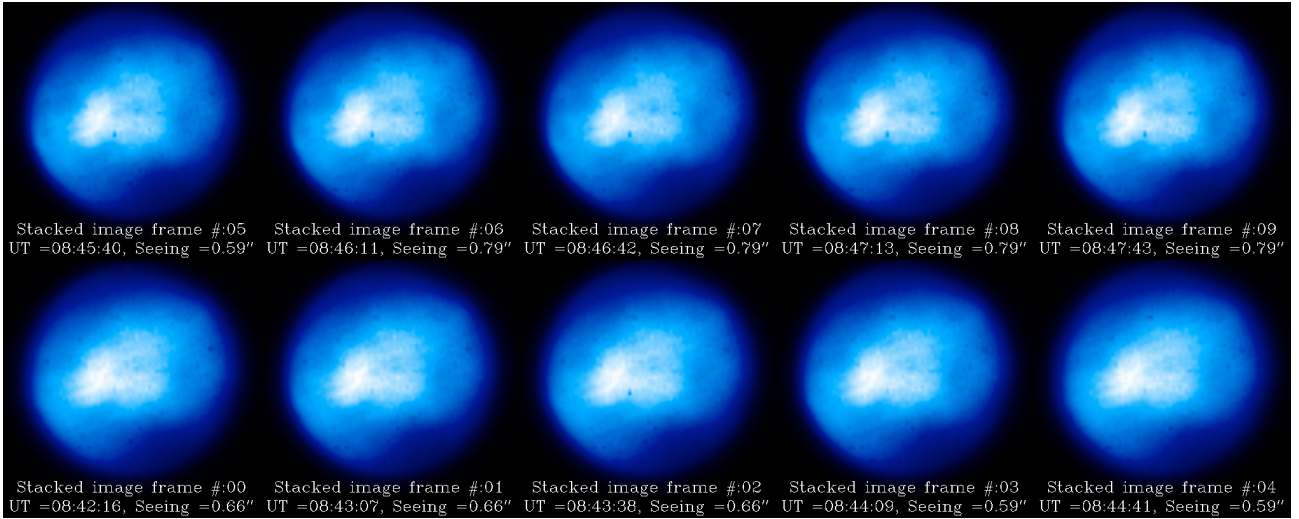

**Figure 5**: K1 images of Europa obtained during the GPI first light run. These images were used to derive a plate scale of 14.21 ± 0.21 mas/pixel for the GPI IFS.

## 3. POSITION ANGLE

The orientation of the instrument on the sky depends both on rotation of the coordinate frame within GPI and the orientation of GPI on the Instrument Support Structure (ISS). Therefore the direction of true north is best measured on sky. For these measurements, we used only those binaries that had concurrent measurements from Keck. The position angles of these binaries are given in Table 1. For GPI data, we used the pipeline primitive "Rotate North Up" to first orient the binaries to the assumed correct north position for GPI.[10] We then calculated their position angle by using the centroids derived in Section 2.2. These angles are listed in Table 4. The difference between the "true" north position from Keck and the position from GPI is the north offset that must be applied to GPI astrometry. This difference is listed in Table 4. Combining all measurements via weighted average gives a final value of -1.00 ± 0.03° for the GPI north offset. Future versions of the GPI data pipeline will incorporate this correction into an updated version of the "Rotate North Up" primitive.

Table 4. GPI log of position angle measurements.

| Object | Date Observed | GPI Measured Position Angle (deg) | Implied North Offset (deg) |
|---|---|---|---|
| HIP 43947B | 14 May 2014 | 261.85 ± 0.02 | -1.03 ± 0.04 |
| HIP 44804A | 14 March 2014 | 307.239 ± 0.002 | -0.92 ± 0.04 |
| " | 14 May 2014 | 307.22 ± 0.19 | -0.90 ± 0.19 |
| HIP 80628 | 14 March 2014 | 44.89 ± 0.12 | -1.27 ± 0.12 |
| " | 11 May 2014 | 44.75 ± 0.08 | -1.13 ± 0.08 |

## 4. DISTORTION

### 4.1 Lab Measurements

In order to obtain astrometry to ~1 mas with GPI, a good model of the IFS distortion is necessary. Previous estimates of distortion for other instruments using laboratory pinhole grids have been limited in their accuracy by unknown uncertainty in grid point spacing, or regular patterns that do not sample important terms like skew.[24] To avoid these issues, we designed grids for GPI such that we could follow the methodology of Anderson & King [25]. In their measurements with the Wide Field Planetary Camera 2 (WFPC2) on the Hubble Space Telescope,[26] they derived a distortion solution self-consistently by using a large number of measurements of a densely populated field at various positions and orientations to measure distortion. In this framework, the true separations of points do not matter – only the ability to uniquely identify which point is which is important. We therefore added additional transparent spots to the square regular pinhole grid described in Section 2.1 so that each spot could be uniquely identified. These additional spots were slightly offset from the others and placed near the edge of the grid. In addition, we also designed a grid with a random pattern on it to simulate a dense star field on the sky – in the case of this grid, the irregular pattern makes the spots easy to identify. A sample observation of the random pattern grid is shown in Figure 6.

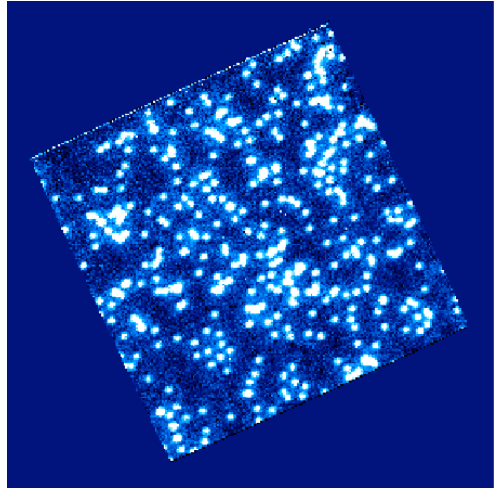

**Figure 6:** Example image of the random pinhole grid designed to simulate a dense star field. This grid was used in conjunction with the square pattern grid shown in Figure 2 to measure the geometric distortion in GPI.

Data were taken for these measurements on 10 December 2012. The pinhole grids were illuminated using an integrating sphere in the GPI telescope simulator as described in Section 2.1. The grids were rotated such that the pattern was sampled through 360° in 11 IFS frames and at various (slight) offsets, all in *H*-band unocculted mode. Data were reduced and transformed into data cubes using the GPI data reduction pipeline. The position of each spot was then measured in all frames in order to conduct a comparative analysis in which all other frames were transformed into the reference frames of a "baseline" using only rotation and offset.[25] The difference between the position of each spot in the "baseline" frame and the transformed frames are averaged to produce positional residuals. These residuals were then averaged for each 20x20 pixel region on the IFS, combining data from both the regular and random grid patterns. A map of the residual vectors is shown in Figure 7 (magnified by a factor of 30).

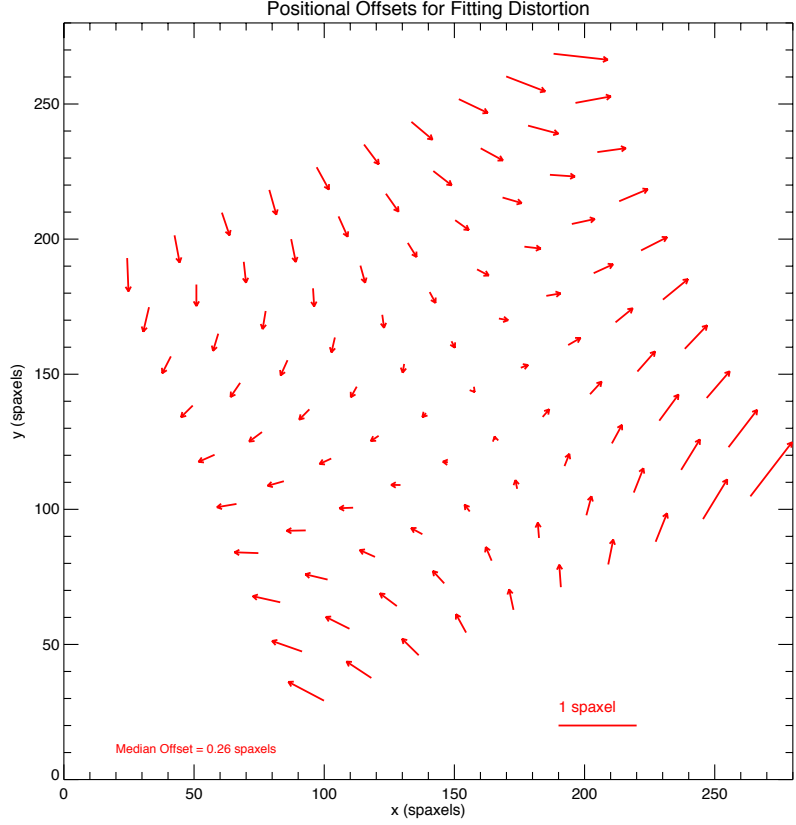

**Figure 7:** Positional residuals across the GPI IFS reconstructed field of view, averaged into 20x20 sections. The median residual size is 0.26 spatial pixels (spaxels). The displacement vectors are magnified by a factor of thirty with respect to the pixel scale.

For our current solution, we opted to model these vectors with a 5$^{th}$ order polynomial with coefficients $[a_n, b_n]$, n=0 → 20, such that:

$x' = a_0 + a_1 x + a_2 y + a_3 x^2 + a_4 xy + a_5 y^2 + a_6 x^3 + a_7 x^2 y + a_8 xy^2 + a_9 y^3 + a_{10} x^4 + a_{11} x^3 y + a_{12} x^2 y^2 + a_{13} xy^3 + a_{14} y^4 + a_{15} x^5 + a_{16} x^4 y + a_{17} x^3 y^2 + a_{18} x^2 y^3 + a_{19} xy^4 + a_{20} y^5$

and

$y' = b_0 + b_1 x + b_2 y + b_3 x^2 + b_4 xy + b_5 y^2 + b_6 x^3 + b_7 x^2 y + b_8 xy^2 + b_9 y^3 + b_{10} x^4 + b_{11} x^3 y + b_{12} x^2 y^2 + b_{13} xy^3 + b_{14} y^4 + b_{15} x^5 + b_{16} x^4 y + b_{17} x^3 y^2 + b_{18} x^2 y^3 + b_{19} xy^4 + b_{20} y^5$

where $x = x_{obs} - 140$ and $y = y_{obs} - 140$.

We experimented with polynomials of different order, and noticed no improvement beyond 5$^{th}$ order. The largest terms in the solution are $a_1$ and $b_1$, which are both approximately 1. This implies that the dominant terms in the GPI distortion are skew terms.[25] The best-fit coefficients are incorporated into the data pipeline as an editable calibration file. The "Correct Distortion" primitive performs a bilinear interpolation of each wavelength channel using the equations above.[10] Work is ongoing modify the primitive such that correcting for distortion preserves flux – however, the scale of the distortion is small enough that the error introduced to the flux is less than other portions of the data pipeline.

Our nominal test for whether the distortion correction was improving our astrometry was to verify that distortion-corrected images had a residual positional offset of less than 0.1 pixels. To perform this test, cubes from the data set were first corrected for distortion, and then an identical analysis was performed on each frame. The new positional residuals were derived, and had a median value of 0.04 pixels. This is well below our requirement, thus implying that our first distortion map is a reasonably good estimate of the true IFS distortion. Figure 8 shows positional residuals in a sample frame before and after distortion correction.

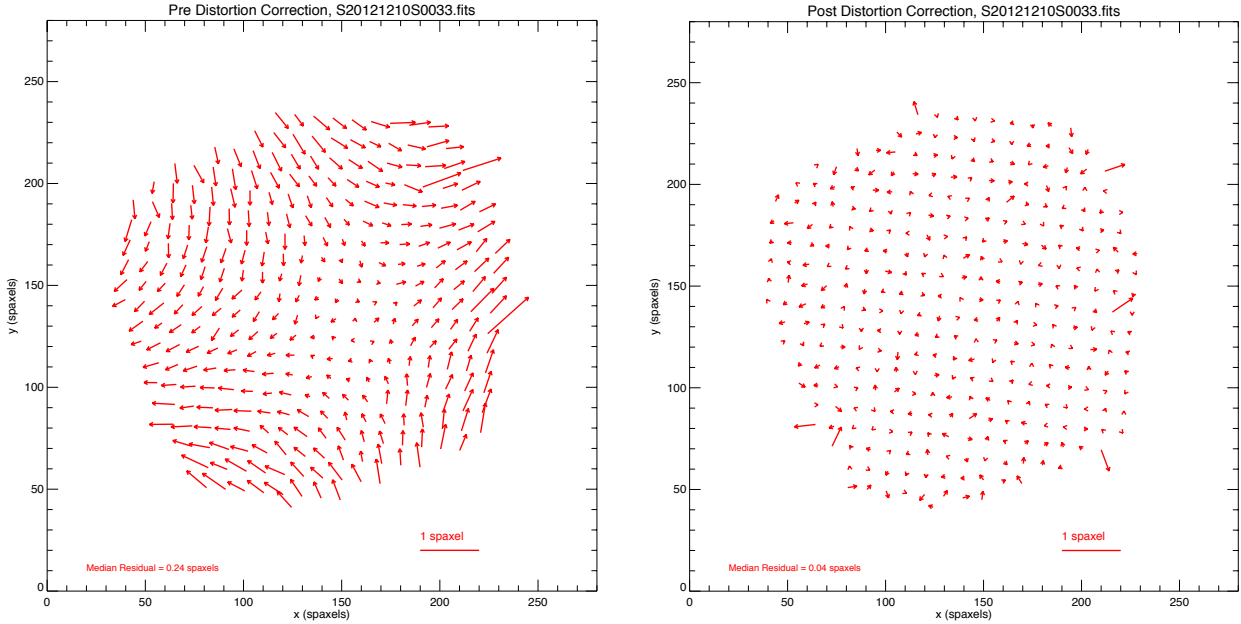

**Figure 8:** Positional residuals for an example IFS wavelength channel pre and post-distortion correction (left and right, respectively). Vectors are magnified by a factor of 30. The median drops from 0.24 pixels to 0.04 pixels when the correction applied. Large vector "outliers" are likely due to poor centroid measurements for an individual spot in one frame.

### 4.2 On-sky Measurements

On-sky measurements to calibrate distortion are necessary to account for any additional affects from the Gemini telescope, which has an estimate geometrical image spread of 0.1 arcseconds over a 1 arcminute field of view.[27] However, these measurements have not yet been completed, as we have not been able to identify a field with a sufficiently bright guide star and stellar density in the GPI field of view to measure distortion properly as was done in the lab. Efforts are ongoing to identify a field for this purpose. In the meantime, we are using the distortion measurements from the laboratory in the data reduction pipeline.

## 5. EXOPLANET SURVEY SIMULATIONS

In order to determine the expected orbital parameter constraints we can obtain during the three-year Gemini Planet Imager Exoplanet Survey (GPIES), a series of Monte Carlo simulations were performed assuming an ideal astrometric precision of 1 mas. We used as input the theoretical planet catalog generated by McBride et al. [28]. This catalog included assumed orbital elements for each planet, which we then used to generate simulated astrometric data. Noise was injected into the data by sampling from a Gaussian distribution centered on the "perfect" relative $x$ and $y$ position with 1-$\sigma$ uncertainties of 1 mas.

The first set of simulations focused on determining the expected eccentricity constraints achievable as a function of number of measurements obtained over the nominal campaign timescale of 3 years. A simplifying assumption was made for all planets that the first epoch of data would be taken in the first semester of the campaign. We then simulated an additional two to eight astrometric measurements, in every case assuming that at least one measurement was taken near the end of the campaign. We then fit orbits to the simulated data using the method described in Ghez et al. [29]. This orbit fitting yields a probability density function (PDF) for each of the six astrometric orbital parameters (period, eccentricity, time of periastron passage, inclination, longitude of the ascending node, and longitude of periastron passage). Taking the PDFs for eccentricity, we calculated the upper and lower limits on this parameter expected for the given number of measurements. The results for a subsample of seven simulated planets are shown in Figure 9. From these simulations, we determined that three measurements, two near the beginning of the campaign and one near the end, would yield the best eccentricity limits with the least expenditure of telescope time. Indeed, somewhat unsurprisingly having the longest time baseline possible has the greatest impact on the parameter distributions.

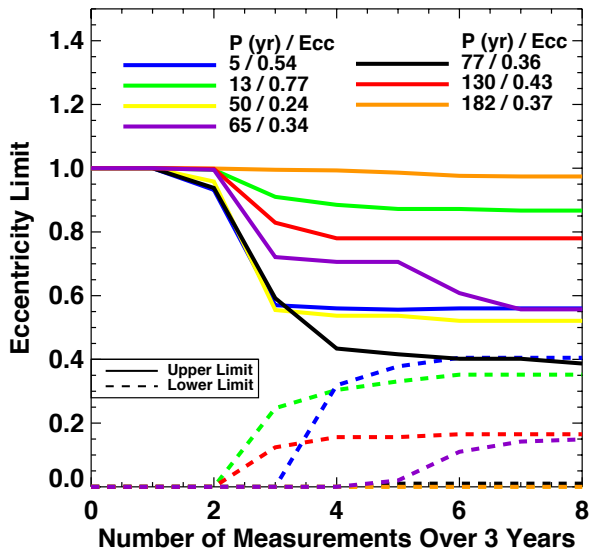

**Figure 9:** Eccentricity limits for a subsample of seven hypothetical GPIES discoveries as a function of the number of measurements over three years. Periods and eccentricities of each planet are given in the legend. Some limits can be placed on all planets except the one with the longest period. Using these results, we determined that three measurements over the lifetime of the campaign is an ideal sampling cadence.

We then considered the resulting PDFs from the three-measurement simulations and how those PDFs could be converted into an overall eccentricity distribution for the sample. We again performed a Monte Carlo simulation, this time in each run sampling an eccentricity from each planet's PDF. This gave us one simulated distribution per iteration. We combined the ensemble of distributions into a series of PDFs in each eccentricity bin of width 0.1. This gives the range of probable eccentricity values for the sample.

The simulations in McBride et al. [28] assumed an eccentricity distribution consistent with that derived for radial velocity planets. In order to determine whether it was possible to distinguish between different underlying distributions, we performed the same simulations, but changed the planet eccentricities to follow two other distributions – an all zero distribution and a thermal distribution. The results of all three simulations are shown in Figure 10. The shaded regions represent the 3-σ range of values per bin. While in a number of bins the distributions are indistinguishable, we find that in the lowest and the highest eccentricity bins, it might be possible to distinguish between the distributions. Future simulations will test other underlying distributions and include more realistic estimates of the astrometric performance of GPI and the timescale between measurement epochs.

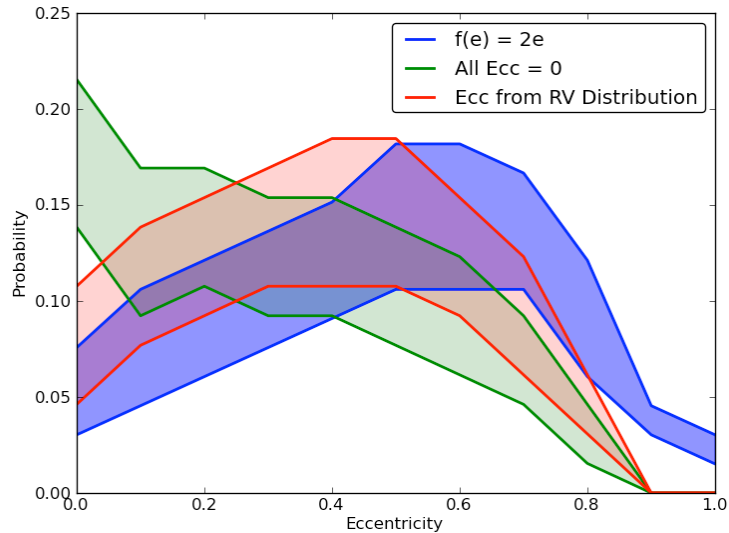

**Figure 10:** The shape of the recovered eccentricity distributions assuming three measurements per planet over a three-year campaign. Three different underlying distributions were considered. The shaded regions represent 3-σ limits per bin. The easiest locations to distinguish between distributions are at *e*=0 and *e* ≥ 0.9. At moderate eccentricities, the distributions are indistinguishable.

## ACKNOWLEDGEMENTS

The Gemini Observatory is operated by the Association of Universities for Research in Astronomy, Inc., under a cooperative agreement with the NSF on behalf of the Gemini partnership: the National Science Foundation (United States), the National Research Council (Canada), CONICYT (Chile), the Australian Research Council (Australia), Ministério da Ciência, Tecnologia e Inovação (Brazil), and Ministerio de Ciencia, Tecnología e Innovación Productiva (Argentina). This publication makes use of data obtained at the W.M. Keck Observatory, which is operated as a scientific partnership among the California Institute of Technology, the University of California and the National Aeronautics and Space Administration. The Observatory was made possible by the generous financial support of the W.M. Keck Foundation. P.K. and J.R.G. thank support from NASA NNX11AD21G, NSF AST-0909188, and the University of California LFRP-118057. Q.M.K is a Dunlap Fellow at the Dunlap Institute for Astronomy & Astrophysics, University of Toronto. The Dunlap Institute is funded through an endowment established by the David Dunlap family and the University of Toronto.

## REFERENCES

[1] Macintosh, B., Graham, J. R., Ingraham, P., Konopacky, Q., Marois, C., Perrin, M., Poyneer, L., Bauman, B., Barman, T., Burrows, A., Cardwell, A., Chilcote, J., De Rosa, R. J., Dillon, D., Doyon, R., Dunn, J., Erikson, D., Fitzgerald, M., Gavel, D., Goodsell, S., Hartung, M., Hibon, P., Kalas, P. G., Larkin, J., Maire, J., Marchis, F., Marley, M., McBride, J., Millar-Blanchaer, M., Morzinski, K., Norton, A., Oppenheimer, B.R., Palmer, D., Patience, J., Pueyo, L., Rantakyro, F., Sadakuni, N., Saddlemyer, L., Savransky, D., Serio, A., Soummer, R., Sivaramakrishnan, A., Song, I., Thomas, S., Wallace, J. K., Wiktorowicz, S., & Wolff, S., "First light of the Gemini Planet Imager," Proceedings of the National Academy of Sciences (2014).

[2] Beuzit, J.-L., Feldt, M., Dohlen, K., Mouillet, D., Puget, P., Wildi, F., Abe, L., Antichi, J., Baruffolo, A., Baudoz, P., Boccaletti, A., Carbillet, M., Charton, J., Claudi, R., Downing, M., Fabron, C., Feautrier, P., Fedrigo, E., Fusco, T., Gach, J.-L., Gratton, R., Henning, T., Hubin, N., Joos, F., Kasper, M., Langlois, M., Lenzen, R., Moutou, C., Pavlov, A., Petit, C., Pragt, J., Rabou, P., Rigal, F., Roelfsema, R., Rousset, G., Saisse, M., Schmid, H.-M., Stadler,

E., Thalmann, C., Turatto, M., Udry, S., Vakili, F., and Waters, R., "SPHERE: a 'planet finder' instrument for the VLT," Proc. SPIE, (2008).

[3] Udry, S. & Santos, N.C. "Statistical Properties of Exoplanets", ARA&A, 45, 397 (2007).

[4] Veras, D., Crepp, J.R., & Ford, E.B., "Formation, Survival, and Detectability of Planets Beyond 100 AU", ApJ, 696, 1600 (2009).

[5] Marois, C., Lafrenière, D, Macintosh, B., & Doyon, R., "Accurate Astrometry and Photometry of Saturated and Coronagraphic Point Spread Functions", ApJ, 647, 612 (2006).

[6] Sivaramakrishnan, A. & Oppenheimer, B.R. "Astrometry and Photometry with Coronagraphs", ApJ, 647, 620 (2006).

[7] Wang, J.J., Rajan, A., Graham, J.R., Savransky, D., Ingraham, P.J., Ward-Duong, K., Patience, J., De Rosa, R.J., Bulger, J., Sivaramakrishnan, A., Perrin, M.D., Thomas, S.J., Sadakuni, N., Greenbaum, A.Z., Pueyo, L., Marois, C., Oppenheimer, B.R., Kalas, P., Cardwell, A., Goodsell, S., Hibon, P., & Rantakyrö, F.T., "Gemini Planet Imager Observational Calibrations VIII: Characterization and Role of Satellite Spots", Proc SPIE 9147 (2014).

[8] Chilcote, J.K., Larkin, J.E., Maire, J., Perrin, M.D., Fitzgerald, M.P., Doyon, R., Thibault, S., Bauman, B., Macintosh, B.A., Graham, J.R., and Saddlemyer, L., "Performance of the integral field spectrograph for the Gemini Planet Imager", Proc. SPIE 8446 (2012).

[9] Macintosh, B.A., Anthony, A., Atwood, J., Barriga, N., Bauman, B., Caputa, K., Chilcote, J., Dillon, D., Doyon, R., Dunn, J., Gavel, D.T., Galvez, R., Goodsell, S.J., Graham, J.R., Hartung, M., Isaacs, J., Kerley, D., Konopacky, Q., Labrie, K., Larkin, J.E., Maire, J., Marois, C., Millar-Blanchaer, M., Nunez, A., Oppenheimer, B.R., Palmer, D.W., Pazder, J., Perrin, M., Poyneer, L.A., Quirez, C., Rantakyro, F., Reshtov, V., Saddlemyer, L., Sadakuni, N., Savransky, D., Sivaramakrishnan, A., Smith, M., Soummer, R., Thomas, S., Wallace, J.K., Weiss, J., & Wiktorowicz, S., "The Gemini Planet Imager: integration and status", Proc. SPIE, 8446 (2012).

[10] Perrin, M., Maire, J., Ingraham, P. J., Savransky, D., Millar-Blanchaer, M., Wolff, S. G., Ruffio, J.-B., Wang, J. J., Draper, Z. H., Sadakuni, N., Marois, C., Rajan, A., Fitzgerald, M. P., Macintosh, B., Graham, J. R., Doyon, R., Larkin, J. E., Chilcote, J. K., Goodsell, S. J., Palmer, D. W., Labrie, K., Beaulieu, M., Rosa, R. J. D., Greenbaum, A. Z., Hartung, M., Hibon, P., Konopacky, Q. M., Lafreniere, D., Lavigne, J.-F., Marchis, F., Patience, J., Pueyo, L. A., Rantakyro, F., Soummer, R., Sivaramakrishnan, A., Thomas, S. J., Ward-Duong, K., and Wiktorowicz, S., "Gemini Planet Imager observational calibrations I: overview of the GPI data reduction pipeline," Proc. SPIE 9147 (2014).

[11] Diolaiti, E., Bendinelli, O., Bonaccini, D., Close, L.M., Currie, D.G., & Parmeggiani, G., "StarFinder: an IDL GUI-based code to analyze crowded fields with isoplanatic correcting PSF fitting", Proc. SPIE 4007 (2000).

[12] Maire, J., Perrin, M. D., Doyon, R., Artigau, E., Dunn, J., Gavel, D. T., Graham, J. R., Lafrenière, D., Larkin, J. E., Lavigne, J.-F., Macintosh, B. A., Marois, C., Oppenheimer, B., Palmer, D. W., Poyneer, L. A., Thibault, S., and Véran, J.-P., "Data reduction pipeline for the Gemini Planet Imager," Proc. SPIE 7735 (2010).

[13] Thomas, S.J., Poyneer, L., de Rosa, R., Macintosh, B., Dillon, D., Wallace, J.K., Palmer, D., Gavel, D., Bauman, B., Saddlemyer, L., & Goodsell, S., "Integration and test of the Gemini Planet Imager", Proc. SPIE 8149 (2011).

[14] Yelda, S., Lu, J.R., Ghez, A.M., Clarkson, W., Anderson, J., Do, T., & Matthews, K., "Improving Galactic Center Astrometry by Reducing the Effects of Geometric Distortion", ApJ, 725, 331 (2010).

[15] Tokovinin, A., Hartung, M., & Hayward, T., "Subsystems in Nearby Solar-type Wide Binaries", AJ, 140, 510 (2010).

[16] De Rosa, R.J., Bulger, J., Patience, J., Leland, B., Macintosh, B., Schneider, A., Song, I., Marois, C., Graham, J.R., Bessell, M., & Doyon, R., "A Volume-limited A-Star (VAST) survey – I. Companions and the unexpected X-ray detection of B6-A7 Stars", MNRAS, 415, 854 (2011).

[17] Raghavan, D., McAlister, H.A., Henry, T.J., Latham, D.W., Marcy, G.W., Mason, B.D., Gies, D.R., White, R.J., & ten Brummelaar, T.A., "A Survey of Stellar Families: Multiplicity of Solar-type Stars", ApJS, 190, 1 (2010).

[18] Metchev, S.A. & Hillenbrand, L.A., "The Palomar/Keck Adaptive Optics Survey of Young Solar Analogs: Evidence for a Universal Companion Mass Function", ApJS, 181, 62 (2009).

[19] Biller, B.A., Liu, M.C., Wahhaj, Z., Nielsen, E.L., Close, L.M., Dupuy, T.J., Hayward, T.L., Burrows, A., Chun, M., Ftaclas, C., Clarke, F., Hartung, M., Males, J., Reid, I.N., Shkolnik, E.L., Skemer, A., Tecza, M., Thatte, N., Alencar, S.H.P., Artymowicz, P., Boss, A., de Gouveia Dal Pino, E., Gregorio-Hetem, J., Ida, S., Kuchner, M.J., Lin, D., & Toomey, D., "The Gemini NICI Planet-finding Campaign: Discovery of a Close Substellar Companion to the Young Debris Disk Star PZ Tel", ApJ, 720, 82 (2010).

[20] Crepp, J.R., Johnson, J.A., Howard, A.W., Marcy, G.W., Gianninas, A., Kilic, M., & Wright, J.T., "The TRENDS High-contrast Imaging Survey. III. A Faint White Dwarf Companion Orbiting HD 114174", ApJ, 774, 1 (2013).

[21] Close, L.M., Males, J.R., Morzinski, K., Kopon, D., Follette, K., Rodigas, T.J., Hinz, P., Wu., Y.-L., Puglisi, A., Esposito, S., Riccardi, A., Pinna, E., Xompero, M., Briguglio, R., Uomoto, A., & Hare, T., "Diffraction-limited Visible Light Images of Orion Trapezium Cluster with the Magellan Adaptive Secondary Adaptive Optics System (MagAO)", ApJ, 774, 94 (2013).

[22] Mugrauer, M., Röll, T., Ginski, C., Vogt, N., Neuhäuser, R., & Schmidt, T.O.B., "New observations of the PZ Tel system, its substellar companion and debris disc", MNRAS, 424, 1714 (2012).

[23] Larkin, J. E., Chilcote, J. K., Aliado, T., Bauman, B. J., Brims, G., Canfield, J. M., Dillon, D., Doyon, R., Dunn, J., Fitzgerald, M. P., Graham, J. R., Goodsell, S., Hartung, M., Ingraham, P. J., Johnson, C. A., Kress, E., Konopacky, Q. M., Macintosh, B. A., Magnone, K. G., Maire, J., McLean, I. S., Palmer, D., Perrin, M. D., Quiroz, C., Sadakuni, N., Saddlemyer, L., Thibault, S., Thomas, S. J., Vallee, P., and Weiss, J. L., "The integral field spectrograph for the gemini planet imager," Proc. SPIE 9147, (2014).

[24] Cameron, P.B, "NIRC2 Geometric Distortion", http://www.astro.caltech.edu/~pbc/AO/distortion.pdf

[25] Anderson, J. & King, I.R., "An Improved Distortion Solution for the Hubble Space Telescope's WFPC2", PASP, 115, 113 (2003).

[26] Gonzaga, S. & Biretta, J., et al., 2010, in HST WFPC2 Data Handbook, v 5.0, ed., Baltimore, STScI

[27] Wynne, C.G. "Field Imaging in Very Large Telescopes", MNRAS, 269, 37 (1994).

[28] McBride, J., Graham, J.R., Macintosh, B., Beckwith, S.V.W., Marois, C., Poyneer, L.A., & Wiktorowicz, S.J., "Experimental Design for the Gemini Planet Imager", PASP, 123, 692 (2011).

[29] Ghez, A.M., Salim, S., Weinberg, N.N., Lu, J.R., Do, T., Dunn, J.K., Matthews, K., Morris, M.R., Yelda, S., Becklin, E.E., Kremenek, T., Milosavljevic, M., & Naiman, J., "Measuring Distance and Properties of the Milky Way's Central Supermassive Black Hole with Stellar Orbits", ApJ, 689, 1044 (2008).